\documentclass[prl,reprint,showpacs,amsmath,amssymb,superscriptaddress,longbibliography]{revtex4-1}
\usepackage{graphicx,bm,color,mathptmx}
\newcommand{\Tr}{\mathop{\mathrm{Tr}} \nolimits}

\begin{document}
\title{Achieving quantum-limited optical resolution}

\author{Martin Pa\'{u}r} 
\affiliation{Department of Optics, Palacky University, 
17. listopadu 12, 771 46 Olomouc, Czech Republic}

\author{Bohumil Stoklasa}
\affiliation{Department of Optics, Palacky University, 
17. listopadu 12, 771 46 Olomouc, Czech Republic}

\author{Zdenek Hradil}
\affiliation{Department of Optics, Palacky University, 
17. listopadu 12, 771 46 Olomouc, Czech Republic}

\author{Luis~L.~S\'anchez-Soto}
\affiliation{Departamento de \'Optica, Facultad de F\'{\i}sica,
 Universidad Complutense, 28040~Madrid,  Spain} 
\affiliation{Max-Planck-Institut f\"ur die Physik des Lichts,
  G\"{u}nther-Scharowsky-Stra{\ss}e 1, Bau 24, 91058 Erlangen,
  Germany} 

\author{Jaroslav Rehacek}
\affiliation{Department of Optics, Palacky University, 
17. listopadu 12, 771 46 Olomouc, Czech Republic}

\begin{abstract}
  We establish a simple method to assess the quantum Fisher
  information required for resolving two incoherent point sources with an
  imaging system. The resulting Cram\'er-Rao bound shows that the
  standard Rayleigh limit can be surpassed by suitable coherent
  measurements.  We explicitly find these optimal strategies and
  present a realization for Gaussian and slit apertures. This involves
  a projection onto the optimal bases that is accomplished with
  digital holographic techniques and is compared with a CCD position
  measurement. Our experimental results unequivocally confirm
  unprecedented sub-Rayleigh precision.
\end{abstract}

\maketitle

Optical resolution is a measure of the ability of an imaging system to
resolve spatial details in a signal. As realized long
ago~\cite{Abbe:1873aa}, this resolution is fundamentally determined by
diffraction, which smears out the spatial distribution of light so
that point sources map onto finite spots at the image plane. This
information is aptly encompassed by the point-spread
function~(PSF)~\cite{Goodman:2004aa}.

The diffraction limit was deemed an unbreakable rule, nicely
epitomized by the time-honored Rayleigh
criterion~\cite{Rayleigh:1879aa}: points can be resolved only if they
are separated by at least the spot size of the PSF of the imaging
system.

The conventional means by which one can circumvent this obstruction
are to reduce the wavelength or to build higher numerical-aperture
optics, thereby making the PSF narrower.  In recent years, though,
several intriguing approaches have emerged that can break this rule
under certain special circumstances~\cite{Dekker:1997aa,Hell:2007aa,
  Kolobov:2007aa,Natsupres:2009aa,Hell:2009aa,Patterson:2010aa,
  Cremer:2013aa}.  Despite their success, these techniques are often
involved and require careful control of the source, which is not
always possible in every application.

Quite recently, a groundbreaking proposal~\cite{Tsang:2015aa,
  Nair:2016aa, Ang:2016aa} has re-examined this question from the
alternative perspective of quantum metrology.  The chief idea is to
use the quantum Fisher information to quantify how well the separation
between two poorly resolved incoherent point sources can be estimated.
The associated quantum Cram\'er-Rao lower bound (qCRLB) gives a fair
bound of the accuracy of that estimation. Surprisingly enough, the
qCRLB maintains a fairly constant value for any separation of the
sources, which implies that the Rayleigh criterion is secondary to the
problem at hand.

In this Letter, we elaborate on this issue presenting quite a
straightforward way of determining the quantum Fisher
information. More importantly, we find the associated optimal
measurement schemes that attain the qCRLB.  We study examples of
Gaussian and sinc PSFs, and implement our new method in a compact and
reliable setup. For distances below the Rayleigh limit, the
uncertainty of this measurement is much less than with direct imaging.

Let us first set the stage for our simplified model. We follow
Lord Rayleigh's lead and assume quasimonochromatic paraxial waves with
one specified polarization and one spatial dimension, $x$ denoting the
image-plane coordinate. To facilitate possible generalizations, we
phrase what follows in a quantum parlance.  A coherent wave of complex
amplitude $U( x )$ can thus be assigned to a ket $| U \rangle $, such
that $U( x )=\langle x | U \rangle$, where $| {x} \rangle$ is a vector
describing a point-like source located at~${x}$.

Moreover, we consider a spatially-invariant unit-magnification imaging
system  characterized by its PSF, which  represents its normalized
intensity response to a point light source.   We shall denote
 this PSF as $I (x) = | \langle x | \psi \rangle |^{2} = |\psi (x)|^{2}$.

 Two incoherent point sources are imaged by that system. For
 simplicity, we consider them to have equal intensities and to be
 located at two unknown points $X_{\pm} = \pm d$ of the object
 plane. The task is to give a sensible estimate of the separation
 $\delta = X_{+} - X_{-}$.  The relevant density matrix, which
 embodies the image-plane modes, can be jotted down as
 \begin{equation}
   \label{eq:densmat}
   \varrho_{d} = \tfrac{1}{2} ( 
   | \psi_{+} \rangle \langle \psi_{+}| +
   | \psi_{-} \rangle \langle \psi_{-}| ) \,, 
 \end{equation}
 where $| \psi_{\pm} \rangle = \exp( \pm i d P) | \psi \rangle$, and
 $P$ is the momentum operator that generates displacements in the $x$
 variable. In the $x$-representation, $\varrho_{d}$ appears as the
 normalized mean intensity profile, which is the image of
 spatially-shifted PSFs; namely,
 $ \varrho_{d} (x) = \tfrac{1}{2} ( |\psi(x-d)|^2 + |\psi(x+d)|^2)$. 
This confirms that the momentum acts as a derivative
 $P = - i \partial_{x}$, much in the same way as in quantum optics.

For points close enough together ($d \ll 1$), which we shall asume
henceforth, a linear expansion gives
\begin{equation}
  \label{eq:psipm}
  | \psi_{\pm} \rangle = \mathcal{N}  \left ( 1 \pm 
    i d P \right ) | \psi \rangle \, ,
\end{equation}
where
$\mathcal{N} = [ 1 + d^{2} \langle \psi | P^{2} | \psi
\rangle]^{-1/2}$
is a normalization constant.  The crucial point is that
$\langle \psi_{-} | \psi_{+} \rangle \neq 0$, so the spatial modes
excited by the two sources are not orthogonal, in general. This
overlap is at the heart of all the difficulties of the problem, for it
implies that the two modes cannot be separated by independent
measurements.

To bypass this problem, we bring in the symmetric and antisymmetric
states
\begin{eqnarray}
  \label{eq:modes}
  | \psi_{s} \rangle & = & C_{s}
   ( | \psi_{+} \rangle +  | \psi_{-} \rangle ) \simeq 
    | \psi \rangle  \, , \nonumber \\
  | \psi_{a} \rangle & = &  C_{a}
    ( | \psi_{+} \rangle -  | \psi_{-} \rangle ) \simeq  
    \frac{P | \psi \rangle}{\sqrt{\langle \psi | P^{2} | \psi \rangle}}  \, ,
\end{eqnarray}
where $C_{a}$ and $C_{s}$ are normalization constants.  When
$\langle \psi | P | \psi \rangle = 0$, these modes are
orthogonal. This happens when; for example, the PSF is inversion
symmetric, which encompasses most of the cases of interest. The modes
in (\ref{eq:modes}) constitute a natural set for writing the density
operator. Actually, in this set $\varrho_{d}$ is diagonal
$ \varrho_{d} | \psi_{j} \rangle = p_{j } |\psi_{j} \rangle$, with
eigenvalues $p_{a} = d^{2} \langle \psi | P^{2} | \psi \rangle$ and
$p_{s} = 1- p_{a}$.

%%%%%%%%%%%%%%%%%%%%%%%%%%%%%%%%
 \begin{figure}[t]
   \centering {\includegraphics[width=\columnwidth]{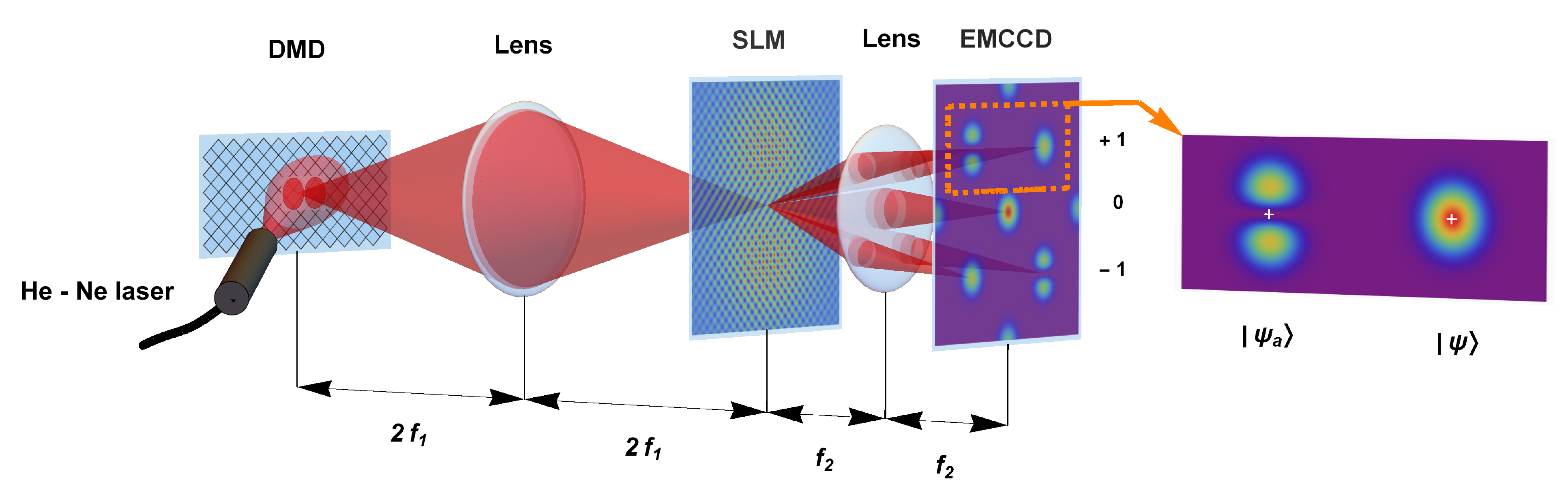}}
   \caption{Schematic diagram of the experimental setup. Two
     incoherent point sources are created with a high frequency
     switched digital micromirror chip (DMD) illuminated with an
     intensity stabilized He-Ne laser. The sources are imaged by a
     low-aperture lens. In the image plane, projection onto different
     modes is performed with a digital hologram created with an
     amplitude spatial light modulator (SLM).  Information about the
     desired projection  is carried by the first-order diffraction spectrum, which is
     mapped by a lens onto a EMCCD camera.}
   \label{fig:setup}
 \end{figure}
%%%%%%%%%%%%%%%%%%%%%%%%%%%%%%%%%%

%%%%%%%%%%%%%%%%%%%%%%%%%%%%%%%%%%
 \begin{figure*}[t]
   \centering
   \includegraphics[width=1.60\columnwidth]{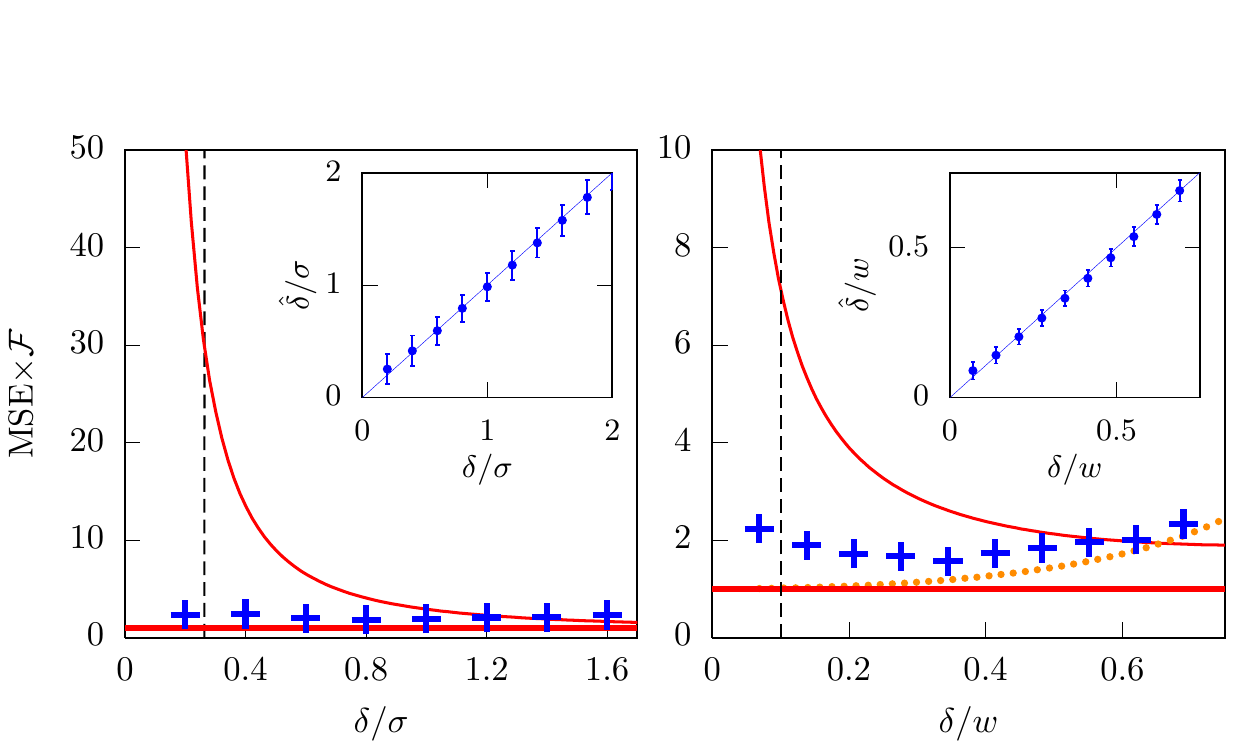}
   \caption{Mean-square error (MSE) of the estimated separation for
     Gaussian (left panel) and sinc (right panel) PSFs. Separations
     are expressed in units of PSF widths $\sigma$ and $w$ and the MSE
     in units of the qCRLB.  The main graph compares the performance
     of our experimental method (blue symbols) with the theoretical
     lower bound for the CCD measurement (thin red curve) and the
     ultimate quantum limit (thick red line). The vertical dotted
     lines delimit the 10\% of the Rayleigh limit for each PSF.  The
     insets show the statistics of the experimental estimates. Mean
     values are plotted in blue dots with standard deviation bars
     around. The true values are inside the standard deviation
     intervals for all separations and the estimator bias is
     negligible.  For the two largest measured separations, the
     experimental MSE nicely follows the classical CRLB calculated for
     the experimentally realized antisymmetric projection (orange
     dots).}
   \label{fig:results}
 \end{figure*}
%%%%%%%%%%%%%%%%%%%%%%%%%%%%%%%%%%%%%%

As we are estimating the separation
$\delta$, the pivotal quantity is the quantum Fisher
information~\cite{Motka:2016aa}. It is  defined as $\mathcal{F} = 
\Tr [ \varrho_{d} L_{d}^{2} ]$, where the symmetric logarithmic
derivative $L_{d}$ is the selfadjoint operator satisfying
$\tfrac{1}{2} (L_{d} \varrho_{d} + \varrho_{d} L_{d} ) = \partial
\varrho_{d} / \partial d$~\cite{Petz:2011aa}. A direct calculation finds that
\begin{equation}
  \label{eq:Fishq}
  \mathcal{F}  = 2  \left [
  \frac{1}{p_{a}}  
  \langle \psi_{a} | 
  \frac{\partial \varrho_{d}}{\partial d} |
  \psi_{a}\rangle  +
  \frac{1}{p_{s}}  
  \langle \psi_{s} | 
  \frac{\partial \varrho_{d}}{\partial d} |
  \psi_{s}\rangle \right ]
 \simeq
  \langle \psi | P^{2} | \psi \rangle \, ,
\end{equation}
and $ \mathcal{F}$ turns out to be independent of $d$.

The qCRLB ensures that the variance of any unbiased estimator
$\hat{\delta}$ of the quantity $\delta$ is then bounded by the
reciprocal of the Fisher information; viz,
\begin{equation}
  \label{eq:qCRLB}
  (\Delta \hat{\delta})^{2} \ge \frac{1}{\mathcal{F}} = 
  \frac{1}{\langle \psi | P^{2} | \psi \rangle} \, .
\end{equation}
As this accuracy remains also constant, considerable improvement can
be obtained if an optimal quantum measurement, saturating
(\ref{eq:qCRLB}), is implemented.

Before we proceed further, we make a pertinent remark. The
classical Fisher information for this problem reads as
\begin{equation}
  \label{eq:Fishcla}
  \mathcal{F}_{\mathrm{cl}}  = \int_{-\infty}^{\infty}
  \frac{1}{\varrho_{d} (x)} \frac{\partial^{2} \varrho_{d}(x)}{\partial d^{2}}
  \, \mathrm{d}x \, .
\end{equation}
Performing again a first-order expansion in $d$,
$\mathcal{F}_{\mathrm{cl}}$ can be  expressed in terms of the PSF
$I(x)$:
\begin{equation}
  \label{eq:Fishclaf}
  \mathcal{F}_{\mathrm{cl}}  \simeq d^{2} \int_{-\infty}^{\infty}
  \frac{[I^{\prime \prime} (x)]^{2}}{I (x)}   \, \mathrm{d}x \, .
\end{equation}
Now, $\mathcal{F}_{\mathrm{cl}} $ goes to zero quadratically as
$d \rightarrow 0$. This means that, in this classical strategy,
detecting intensity at the image plane is progressively worse at
estimating the separation for closer sources, to the point that the
classical CRLB diverges at $d=0$. This divergent behavior has been
termed the \emph{Rayleigh's curse}~\cite{Tsang:2015aa}.
In other words, there is much more information available about  
the separation of the sources in the phase of the field than in the
intensity alone.   

From our previous discussion, it is clear that
the projectors $\Pi_{j} = |\psi_{j} \rangle \langle \psi_{j}|$
($j = a,s$) comprise the optimal measurements of the parameter $d$.
Notice that in Eq.~(\ref{eq:Fishq}), the antisymmetric mode $p_{a}$
gives the leading contribution and thus most useful information can be
extracted from the $\Pi_{a}$ channel. As a consequence, the wave
function of the optimal measurement becomes
\begin{equation}
  \label{eq:opt}
  \psi_{\mathrm{opt}} (x) = \langle x | \psi_{a}\rangle = 
\frac{\psi^{\prime} (x)}{\sqrt{\mathcal{F} (x)}}
\end{equation}
where the $x$ representation of the quantum Fisher information 
is
\begin{equation}
  \label{eq:FishQx}
  \mathcal{F} (x) = \langle \psi | P^{2}| \psi \rangle = 
\int_{- \infty}^{\infty} [ \psi^{\prime} (x)]^{2} \, \mathrm{d}x \,.  
\end{equation}
Let us consider two relevant examples of PSFs;
the Gaussian and the sinc: 
\begin{equation}
  \label{eq:exPSF}
  \psi^{G} (x) =  \frac{1}
{(2 \pi \sigma^2)^{\frac{1}{4}}} 
\exp{ \left ( - \frac{x^2}{4 \sigma^2} \right )} \,,
\quad 
 \psi^{s} (x) = \frac{1}{\sqrt{w}} \mathrm{sinc}\left ( \frac{\pi x}{w} \right )  \, ,
\end{equation}
where $\sigma$ and $w$ are effective widths that depend on the
wavelength. From Eq.~(\ref{eq:FishQx}) it is straightforward to obtain
the quantum Fisher information for these two cases: $1/(4 \sigma^{2})$
and $\pi^{2} / (3w^{2})$, respectively. The optimal measurements are
then
\begin{eqnarray}
  \label{eq:optproj}
 \psi_{\mathrm{opt}}^{G} (x) & = &  
 \frac{1}{(2 \pi)^{\frac{1}{4}}\sigma^{\frac{3}{2}}} 
 x \, \exp{ \left ( - \frac{x^2}{4 \sigma^2} \right )}  \,, \nonumber \\
  \psi_{\mathrm{opt}}^{S} (x) & = & 
 \sqrt{3} 
\left [ 
 \frac{w^{\frac{1}{2}}}{\pi x}   \cos\left ( \frac{\pi x}{w}  \right )
- \frac{w^{\frac{3}{2}}}{\pi^{2} x^{2}}    \sin \left ( \frac{\pi x}{w}  \right )  
\right ] \, .
\end{eqnarray}

To project on these functions, one needs to separate the image-plane
field in terms of the desired spatial modes. This has been implemented
in our laboratory with the setup sketched in Fig.~\ref{fig:setup}.
Two incoherent point-like sources were generated by a Digital Light
Projector (DLP) Lightcrafter evaluation module (Texas Instruments),
which uses a digital micromirror chip (DMD) with square micromirrors
of 7.6~$\mu$m size each. This allows for a precise control of the
points separation by individually addressing two particular
micromirrors. The DMD chip was illuminated by an intensity-stabilized
He-Ne laser equipped with a beam expander to get a sufficiently
uniform beam. The spatial incoherence is ensured by switching between
the two object points, so that only one was ON at a time, keeping the
switching time well below the detector time resolution.

The two point sources were imaged by a low numerical-aperture lens and
shaped by an aperture placed behind the lens.  A circular diaphragm
produced Airy rings, but these are well approximated by a Gaussian
PSF.  The sinc PSF was obtained by inserting a squared slit.  We
experimentally measured the values $\sigma =0.05$~mm and $w
=0.15$~mm. The Rayleigh criterion for these values are $2.635 \sigma$ 
and $w$, respectively. The two-point separations $\delta$ were varied 
in steps of $0.01$~mm, which corresponds to steps $0.2 \sigma$ for the 
Gaussian and $0.067 w $ for the sinc. The smallest separations
attained  are 13 times smaller than the Rayleigh limit for the Gaussian and 10
times for the sinc.

The projection onto any basis is performed with a spatial light
modulator (CRL OPTO) in the amplitude mode.  We prepare a hologram at
the image plane produced as an interference between a 
tilted reference plane wave and the desired projection function
$\psi_{\mathrm{opt}}$.  When this is illuminated by the two-point
source,  the intensity in the propagation direction of the reference
wave is
\begin{equation}
  \left  | \int^{\infty}_{-\infty} 
    \psi_{\mathrm{opt}}^{\ast} (x) \psi (x+ d ) \mathrm{d}x \right |^{2}+
  \left  | \int^{\infty}_{-\infty} 
    \psi_{\mathrm{opt}}^{\ast}  (x) \psi (x- d ) \mathrm{d}x \right |^{2} \, . 
\end{equation}
Different projections can be obtained with different
reference waves.

To prepare the hologram, nominal PSF parameters ($\sigma$ and $w$)
were measured in advance.  For the Gaussian PSF, we prepared
projection on both the zeroth- and first-order Hermite-Gaussian modes.
The measurement of the zeroth-order mode is used to assess the total
number of photons in each measurement run.  For the sinc, the image
was also projected on the PSF itself and its first spatial derivative.

The desired projection is carried by the first diffraction order. To
get the information, the signal is Fourier transformed by a
short-focal lens and detected by a cooled electron-multiplying CCD
(EMCCD) camera (Raptor Photonics) working in the linear mode with 
on-chip gain to suppress the effects of read-out noise and dark noise. As
sketched in Fig.~\ref{fig:setup}, the outcome of a measurement
consists of two photon counts detected from the Fourier spectrum
points representing spatial frequencies connected with the reference
waves. This data carries information about the separation of the two
incoherent point sources.

The excess noise of EMCCD gain $g$ constitutes a remaining noise
source. The numbers of photons $n_0$ and $n_a$ detected in the PSF
$|\psi\rangle$ and antisymmetric (optimal) modes $|\psi_a\rangle$,
respectively, was determined by using the EMCCCD pixel capacity and
$g$. The relative frequency of measuring the antisymmetric projection
was calculated as $f_a=n_{a}/(n_{0}+n_{a})$, the denominator
$n_{0}+n_{a}$ being roughly the total number of detected photons. The
estimator of the separation is then obtained by solving the relation
$f_a=\langle \psi_a| \varrho_d|\psi_a\rangle$ for $d$. In doing so we
make no assumption about the smallness of $\delta$, which helps to
produce unbiased estimates of larger separations.

To determine estimator characteristics, 500 measurements for each
separation were carried out.  Results are sumarized on the
Fig.~\ref{fig:results}. The optimal method overcomes the direct position
measurement for small and moderate separations. For the
Gaussian PSF (left panel) and the smallest separation $0.2 \sigma$, the
experimental mean squared estimator (MSE) is 2.35$\times$qCRLB; i.e., more
than 20 times smaller than the error of the position measurement
($51.2 \times$qCRLB).  For the sinc, the experimental MSE is $2.23\times$qCRLB
for the smallest separation, which is $4.5$ times lower than the error
of the position measurement ($10.1 \times$qCRLB). We mention in passing
that the optimal measurement of the sinc PSF is more challenging due 
to very fast oscillations of the PSF derivative. Nevertheless, the
results are quite satisfactory good and in complete agreement with the theory. 

In summary, we have developed and demonstrated a simple technique that
surpasses traditional imaging in its ability to resolve two closely
spaced point sources.  The method does not require any exotic
illumination and is applicable to classical incoherent sources.  Much
in the spirit of the original proposal~\cite{Tsang:2015aa,
  Nair:2016aa,Ang:2016aa}, our results stress that diffraction
resolution limits are not a fundamental constraint but, instead, the
consequence of traditional imaging techniques discarding the phase
information present in the light. 

Moreover, our treatment also suggests other directions of research.
Whereas the point source represents a natural unit for image
processing (upon which hinges the very definition of PSF), other
``quantum units'' can be further expanded and processed in a similar
way. Optimal detection can then be tailored to suit the desired
target. We have shown this for two particular cases of
projections. This clearly provides a novel and not yet explored avenue
for image processing protocols.  We firmly believe that this approach
will have a broad range of applications in the near future.

Note: While preparing this manuscript, we came to realize that similar
conclusions, although with different techniques, were being reached by Sheng
\emph{et al}~\cite{Sheng:2016aa}, Yang \emph{et
  al}~\cite{Yang:2016aa}, and Tham \emph{et al}~\cite{Tham:2016aa}.

\section*{Funding Information}
We acknowledge financial support from the Technology Agency of the
Czech Republic (Grant TE01020229), the Grant Agency of the Czech
Republic (Grant No. 15-03194S), the IGA Project of the Palack{\'y}
University (Grant No. IGA PrF 2016-005) and the Spanish MINECO (Grant
FIS2015-67963-P).

\section*{Acknowledgments}
We thank Gerd Leuchs, Robert Boyd, Juan J. Monz\'on, Olivia Di Matteo, 
and Matthew Foreman for valuable discussions and comments.

% Bibliography
%\bibliography{Resolution}

\end{document}